\documentclass{emulateapj}
\usepackage{apjfonts}
\usepackage{lscape}

\slugcomment{Accepted for ApJ Letters}
\shorttitle{SDSS J125637-022452}
\shortauthors{}
\newcommand{\nc}{\newcommand}         
\nc{\kms}{\,${\rm km\,s}^{-1}$\,} 
\nc{\htwo}{\,${\rm H}_{2}$\,} 
\nc{\tio}{\,${\rm TiO}$\,} 
\nc{\sdss}{{\rm SDSS~125637$-$0224}\,} 
\nc{\feh}{\,${\rm Fe/H}$\,} 
\begin{document}


\title{SDSS J125637-022452: a high proper motion L subdwarf}

\author{T. Sivarani\altaffilmark{1,3}, S. L\'epine\altaffilmark{2},
  A. K. Kembhavi\altaffilmark{3}, \& J. Gupchup\altaffilmark{3,4}}

\altaffiltext{1}{Department of Astronomy, University of Florida,
 211 Bryant Space Science Center, Gainesville, FL, 32611-2055, USA}

\altaffiltext{2}{Department of Astrophysics, Division of Physical Sciences,
American Museum of Natural History, Central Park West at 79th Street,
New York, NY 10024, USA}
\altaffiltext{3}{Inter University Center for Astronomy and Astrophysics (IUCAA),
Pune University Campus,Pune 411 007, India}
 
\altaffiltext{4}{ Department of Computer Science, Johns Hopkins University, 3400 N. Charles street.
Baltimore MD 21218, USA}

\email{}


\begin{abstract}
We report the discovery of a high proper motion L subdwarf ($\mu
=0.617\arcsec$ yr$^{-1}$) in the Sloan Digital Sky Survey spectral
database. The optical spectrum from the star SDSS J125637-022452 has
mixed spectral features of both late-M spectral subtype (strong TiO
and CaH at 7000\AA) and mid-L spectral subtype (strong wings of KI at
7700\AA, CrH and FeH), which is interpreted as the signature of a very
low-mass, metal-poor star (ultra-cool subdwarf) of spectral type
sdL. The near infrared (NIR) $(J-K_{s}$) colors from 2MASS shows the
object to be significantly bluer compared to normal L dwarfs, which is
probably due a strong collision induced absorption (CIA) due to \htwo
molecule. This is consistent with the idea that CIA from \htwo is more
pronounced at low metallicities. Proper motion and radial velocity
measurements also indicate that the star is kinematically ``hot'' and
probably associated with the Galactic halo population.
\end{abstract}

\keywords{Galaxy: halo \--- stars: low mass, brown dwarf, subdwarf
  \--- star: individual \objectname{SDSS J125637-022452}}

\section{Introduction}

Ultra-cool dwarfs, low-mass objects of low temperature extending
beyond the classical main sequence, have been identified in
significant numbers from recent large optical and near infrared
surveys, such as the Deep Near Infrared Survey \citep{Eetal97}, the
Sloan Digital Sky Survey \citep{Yetal00}, and the Two Micron All-Sky
Survey \citet{Cetal03}. Two new spectral type (L,T) have been added
to classify those extremely cool objects, and these are now widely in
use, with several hundred L and T dwarfs classified to date
\citep{kirk1999,kirk2000,geballe2002,hawley2002,geballe2002,burgasser2003}.
Most stars classified as L and T dwarfs are relatively metal-rich, and
associated with the Galactic disk population.

One also expects the Solar neighborhood to be host to ultra-cool
members of the Galactic halo (Population II). However Pop II stars are
rare in the vicinity of the Sun, where they account for roughly one
out of every 200 stars. Conversely, one expects ultracool {\em
  subdwarfs} (sdL, sdT) to be equally rare. In any case, old
metal-poor stars and brown dwarfs are expected to display a distinct
spectral signature, making their identification straightforward. In
stars of spectral type M, metal depletion is known to result in a
weakening of metal oxide bands, usually prominent in M stars
\citep{kuiper}. M subdwarfs are thus organized following distinct
classification sequences as subdwarfs (sdM), extreme subdwarfs (esdM),
and ultrasubdwarfs (usdM), depending of the magnitude of
metal-depletion effects in their spectra
\citep{gizis1997,lepine2007}. Note that spectroscopically confirmed M
subdwarfs number only in the hundreds \citet{lepine2007}, compared
with the tens of thousands of stars now classified as M dwarfs.

Very few subdwarfs of spectral subtype sdM7 or later (``ultra-cool
subdwarfs'') have been identified to date. Most have been discovered
in follow-up spectroscopic surveys of faint stars with very large proper
motions \citep{lepine2002,lepine2003aja}, others from the massive
Sloan Digital Sky Survey spectroscopic database
\citep{west,LeSc2008}. Extending the M subdwarf sequence to subtypes
later than sdM7/esdM7/usdM7 has been straightforward as the metal-poor
stars display the same weakening of the TiO bands as document for
earlier subtypes. More challenging has been the identification of
metal-poor stars beyond the spectral type M, in the range of surface
temperature characteristic of the L and T stars, and designated as L
and T subdwarfs (sdL, sdT).

\begin{figure}
\epsscale{2.00}
\plotone{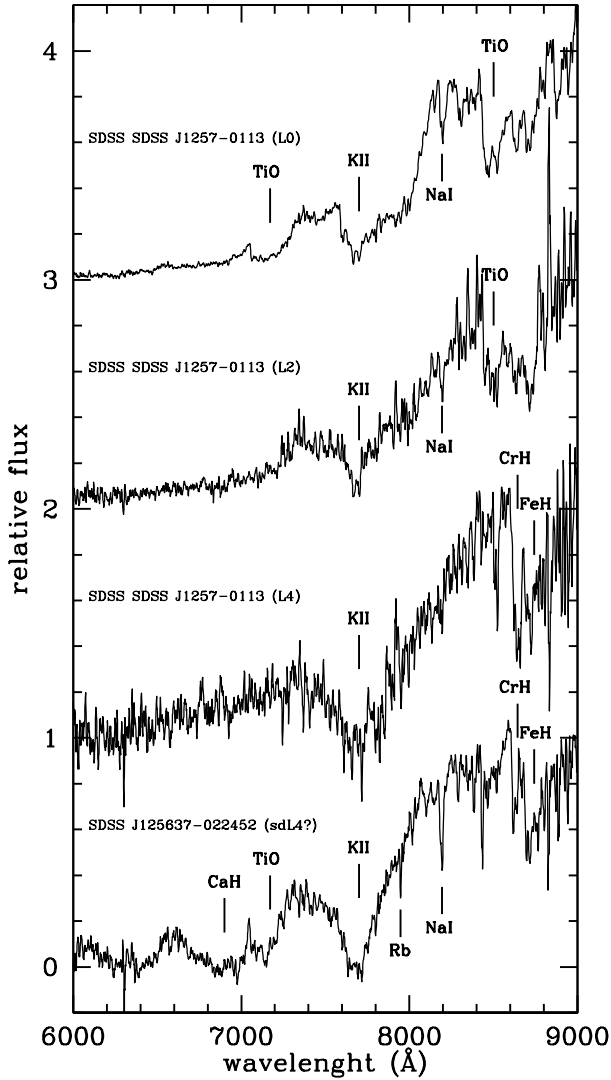}
\caption{Optical SDSS spectrum of the candidate L subdwarf
  SDSS~J125637.1$-$022452 (bottom), compared with spectra of
  early-type L dwarfs, also from SDSS. The deep and broadened KI
  7700\AA\ doublet, is similar to that of an L4 dwarf. However,
  SDSS~J1256 also displays well defined CaH and TiO molecular bands at
  7000\AA, typically seen only in M dwarfs. Elsewhere the hydride
  bands (FeH, CrH) are strong while the oxide (TiO, VO) are weak or
  absent, which is a defining feature of the metal-poor M
  subdwarfs. This suggests that SDSS~J1256 should be classified as a
  subdwarf of type L.}
\label{spec}
\end{figure}

The first star to be unambiguously identification as a subdwarf
spectral type L is 2MASS 0532+8246 \citep{burgasser2003a}. The optical
spectrum had mixed spectral features corresponding to early and late L
spectral types. The NIR spectrum have strong collision induced
absorption (CIA) due to \htwo molecules, giving a blue color similar
to a T dwarf. Another star with similar features (2MASS~1626+3925) was
discovered by \citet{burgasser2004} and also tentatively classified as
sdL. A third object, the star LHS~1610-0040, was initially claimed to
be an early-type subdwarf \citet{lepine2003apjlb}, but further
analysis has failed to substantiate the claim \citep{RB06,C06}; the
star is now believed to be a peculiar late-type dwarf, possibly
displaying anomalous metal abundances \citep{Detal08}.

In this paper we present the discovery of another object with spectral
characteristic consistent with a metal-poor ``L subdwarf''. The star
\sdss was identified from the Sloan Digital Sky Survey spectroscopic
database, and is found to be significantly cooler than all known M
subdwarfs, but warmer than 2MASS~0532+8246 and 2MASS~1626+3925. We
examine the spectral characteristics and kinematics of the star.

%
%
%

\section{Search and identification}

The Sloan Digital Sky Survey obtains spectra from a variety of objects
based on various color and magnitude selection cuts
\citep{stoughton02}. The survey is not complete in most of the star
categories, as a limited number of fibers (640) are used in each of
the SDSS fields, and stellar targets are assigned only after the
primary categories (QSOs, galaxies). The spectra cover the full
3300\AA-9500\AA\ wavelength range, which includes the main molecular
features used to identify cool dwarfs and subdwarfs.

The SDSS second data release (DR2) listed 13,379 spectra of sources
identified as cool and ultra-cool stars (spectral subtype M and
later). The DR2 covered a total survey area of 2627 square degrees or a
little over 6.5\% of the sky. In an attempt to detect ultra-cool L
subdwarfs from this sample, we have systematically examined the spectra
form all stellar sources with very red optical-to-infrared
color. First, we identified all possible counterparts to the 13,379
late-type stars in the 2MASS All-Sky catalog of point source
\citep{Cetal03}. Then we assembled spectra of all the stars with
magnitude $r>$18 and color $(r-K_{s})>$6.0, which eliminating from the
sample most objects with spectral subtypes M6 or earlier. We visually
inspected all the spectra in search of any star with a peculiar
spectral energy distribution. All spectra were found to be consistent
with either late-type M dwarfs or L dwarfs, except for only one which
clearly stood out from the group: the spectrum of the star \sdss.

Sloan photometry shows \sdss to be very faint in the optical, but it
has relatively bright counterparts in both the 2MASS and DENIS
infrared catalogs; the object is clearly very red. It is undetected in
the Digital Sky Survey blue (IIIaJ) and red plates (IIIaF), but has a
counterpart on the infrared (IVn) plates \--- and is thus registered
in the SuperCOSMOS Sky Archive (SSA). Data on this unusual object are
recorded in Table 1. 

\section{Classification and spectral features}

The very red spectrum of \sdss is displayed in Figure 1. The star
shows many spectral features typical of late-M and L dwarfs,
which confirms that it is a very cool object and not a background star
affected by reddening. The dominant feature is a deep K~I doublet at
7700\AA, with strong pressure broadened wings, similar to what is
observed in mid-type L dwarfs \citep{kirk1999,geballe2002}. The
spectrum also displays strong bands of CrH and FeH at 8600\AA\ and
atomic lines of RbI, all typically observed in L
dwarfs. Paradoxically, \sdss also displays well-defined bands of CaH
and TiO around 7050\AA, which are usually observed in M dwarfs and are
normally absent in subtypes later than M9, due to the condensation of
oxides into dust grains. Indeed redward of 7500\AA\ the spectrum is
strongly reminiscent of a late-type M dwarf.

Overall, the spectrum does not fit within the standard M/L
dwarf classification scheme \citep{kirk1999,kirk2000}, and appears to
be a hybrid of M-type and L-type spectral features. However, the
spectrum is strikingly similar to the optical spectrum of the ``sdL''
star 2MASS~1626+3925 \citep{gizis2006,burgasser2007}, with prominent
bands of TiO and CaH redward of 7500\AA\ in what looks in all other
respect like an L dwarf. The lingering presence of TiO bands in those
ultra-cool objects is interpreted as the signature of a metal-poor
atmosphere in which dust formation is inefficient, and which maintains
metal oxides in gaseous form even at very low temperature.

In any case, \sdss is too cool to be classified an M subdwarf.
With $(r-z)=4.15$, \sdss is significantly redder in the optical than
the coolest known M subdwarfs \citep{LeSc2008}. But again paradoxically,
the optical-to-infrared colors are significantly bluer than in field L
dwarfs, and the $(i-J)=3.31$ is more in line with subtype M6-M7
\citep{hawley2002}. Furthermore, the infrared colors of \sdss are unusual:
with $(J-K_{s}) = $0.66, \sdss is significantly bluer that any known field L
dwarfs, which all have $(J-K_{s})>1.0$ \citep{hawley2002}. The same blue
$(J-K_{s})$ color is observed in the L subdwarf 2MASS~0532+82
\citep{burgasser2003a}, which further suggest they are of a similar
class. In 2MASS~0532$+$82 the blue IR color is found to be due to a
strong collision induced absorption (CIA) band from \htwo. Such
unusually strong CIA is also suggested to be a consequence of
low-metallicity \citet{borysow}, and results in a significant
redistribution of the flux in the optical. All in all, the photometry
of \sdss also supports the idea that it a metal-depleted, ultra-cool
dwarf. 

\begin{deluxetable}{lll}
\tabletypesize{\scriptsize}
\tablecolumns{3}
\tablewidth{225pt}
\tablecaption{Data: SDSS 125637-022452}
\tablehead{Datum & Value & Source}
\startdata
R.A.   &  12$^{h}$ 56$^{m}$ 37.1$^{s}$  & SDSS-DR6 \\
Decl.  & -02\degr 24\arcmin 52.5\arcsec & SDSS-DR6   \\ 
$\mu_{R.A.}$  &$-470\pm64$ mas yr$^{-1}$&\\
$\mu_{Decl.}$ &$-378\pm64$ mas yr$^{-1}$&\\
u    & 24.74$\pm$1.18  mag\tablenotemark{1}    & SDSS-DR6 \\
g    & 23.71$\pm$0.37  mag    & SDSS-DR6 \\
r    & 21.82$\pm$0.11  mag    & SDSS-DR6\\
i    & 19.41$\pm$0.02  mag    & SDSS-DR6\\ 
z    & 17.71$\pm$0.02  mag    & SDSS-DR6\\
extinction\_r & 0.06 mag &\citet{Sch98} \\
I    & 18.02  mag    & SSA-DSS 2\tablenotemark{2}\\
I    & 18.36  mag    & SSA-DSS 2\tablenotemark{2}\\
J    & 16.10$\pm$0.11  mag    & 2MASS\\  
H    & 15.79$\pm$0.15  mag    & 2MASS\\
K$_{s}$    & 15.44:  mag    & 2MASS\\
RV   & -90$\pm$40\kms &SDSS-DR2\\
Distance  & $42_{-21}^{+37}$ pc & \\
U    & -17\kms    &\\
V    & -143\kms   &\\
W    & +43\kms   &\\
\enddata
\tablenotetext{1}{psf magnitude}
\tablenotetext{2}{I band magnitude from second epoch DSS plates from
  SuperCOSMOS Sky Archive (SSA)}
\end{deluxetable}

We found an unusually strong FeH bandhead at 8692\AA, relative to CrH
at 8611\AA. \citet{kirk1999} found that the CrH band peaks at spectral
type L7 and the FeH band peaks at L2-L4, consistent with a
condensation temperature for CrH about few hundred degrees cooler than
for FeH. However CrH and FeH are equally strong at L0, and CrH/FeH just
marginally increase at L2 and the bands become equal again at L4. Then
CrH/FeH increase again, but around L6 both FeH and CrH start to grow
weaker. In SDSS~J125637$-$0224, for the first time, we see FeH
stronger than CrH. In 2MASS~0532$+$8246, the L subdwarf identified by
\citet{burgasser2003a}, CrH and FeH are of equal strength; however one
naturally expects a stronger CrH band relative to FeH in this very
cool object ($\sim$L7), as observed also in DENIS~0205$-$1159AB
\citep{burgasser2003}. The most likely explanation for the larger FeH
in \sdss is that it reflects a low relative abundance of Cr to Fe,
compared with solar metallicity L dwarfs. Indeed, in abundance analyses
of halo stars of low metallicities ([Fe/H]$< -$1.0), Cr has been found
to be under-abundant by about 0.4 $dex$ relative to Fe
\citep{mcwilliam1995,first5}.

With all evidence pointing to an ultra-cool subdwarf, and based on the
spectral energy distribution in the optical and the breadth of the K~I
doublet, we classify \sdss as a mid-type L-subdwarf (tentatively
``sdL4'') although complete sequences of such objects would be
required to formally establish spectral subtypes. 

\section{Distance and kinematics}

\begin{deluxetable}{lrrr}
\tabletypesize{\scriptsize}
\tablecolumns{4}
\tablewidth{225pt}
\tablecaption{Observation details}
\tablehead{Date &  RA & DEC & Source}
\startdata
14/03/1987& 12 56 37.60  &-02 24 48.0 & SERC I-UKST\tablenotemark{1}\\
15/05/1996& 12 56 37.27  &-02 24 50.7 & DENIS \\
31/05/1997& 12 56 37.22  &-02 24 50.3 & POSS-II N\tablenotemark{2}\\
11/03/1999& 12 56 37.14  &-02 24 51.6 & UKST\tablenotemark{3}\\
11/02/1999& 12 56 37.17  &-02 24 52.2 & 2MASS  \\
30/05/2000& 12 56 37.13  &-02 24 52.4 & SDSS-DR2 \\
\enddata
\tablenotetext{1}{SERC I, deep I plates made by the Second Epoch
  Survey of the southern sky in I band taken with UK Schmidt
  Telescope(UKST) and digitized by SuperCOSMOS digitizing Machine.}
\tablenotetext{2}{Second Epoch Survey of the northern sky in I band by
  the Palomar Observatory Sky Survey (POSS II N) digitized by Space
  Telescope Science Institute Digital Sky Survey.}
\tablenotetext{3}{Second Epoch Survey of the southern sky in I band
  taken with UKST and digitized by SuperCOSMOS digitizing
  Machine.}
\end{deluxetable}


\begin{figure}[t]
\epsscale{1.0}
\plotone{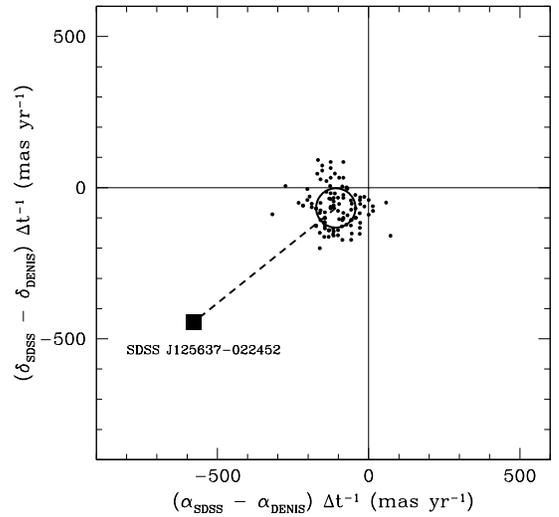}
\caption{Offset between quoted positions in the SDSS catalog
  (epoch=2000.1) and DENIS catalog (epoch=1996.3) for \sdss and 110
  stars within $10\arcmin$ of its position. The significant offset
  of \sdss \ between the two epochs is consistent with a large proper
  motion [$\mu_{\alpha},\mu_{\delta}$] = [$-470\pm64,-378\pm64$] $mas$
  yr$^{-1}$ relative to the background stars (dashed line).}
\label{ppm}
\end{figure}

%
%
%
%

The DENIS and SDSS images on which \sdss is identified are separated
by $\simeq3.7$yr, which is sufficient to extract a proper motion. We
compared the DENIS and SDSS positions of all the stars within
$10\arcmin$ of \sdss to determine any offset between the DENIS and
SDSS coordinate systems, and to estimate the astrometric errors. For
the background stars, we found mean offset consistent with a shift of
-108 mas yr$^{-1}$ in R.A. and -65 mas yr$^{-1}$ in Decl. between the
two epochs, with a 1-$\sigma$ dispersion of 64 mas yr$^{-1}$ in both
coordinates. For \sdss \ we find a significantly larger offset which,
after correcting for the mean offset, indicates a proper motion
relative to background stars of [$\mu_{\alpha},\mu_{\delta}$] =
[$-470\pm64,-378\pm64$] $mas$ yr$^{-1}$ (Figure 2).

Radial velocity for \sdss is obtained directly from the SDSS pipeline,
where it is measured by cross-correlation with a template from a star
of similar spectral energy distribution (most likely an L dwarf), and
is reduced to the heliocentric frame after correcting for Earth's
motion. The quoted radial velocity is 90$\pm$40\kms. 
We have verified this value by cross-correlating the spectrum with L3 and L5
templates \citep{kirk2000}, and obtained 127\kms and 109\kms respectively.
In both cases the cross correlation width is $\sim 40 $\kms.

Without any parallax measurement, any distance estimate for \sdss
should be considered with caution. Because of its unusual spectral
energy distribution, it is unclear whether any of the color-magnitude
relationships defined for M and L dwarfs should apply. Assuming \sdss
\ to be analogous to a mid-type L subdwarf, we apply the
color-magnitude relationships of \citep{dahn} and
\citep{hawley2002}. For an L4 dwarf, the typical absolute J
magnitude is $M_J\simeq13.0$, which we assume to be a reasonable
estimate. In any case, it seems unlikely that \sdss \ should be more
luminous than a late-type M dwarf ($M_J\simeq11.5$), and it should be
somewhat more luminous than a late-type L dwarf
($M_J\approx14.5$). Based on these values, we estimate a spectroscopic
distance of about 42 pc, with an actual distance possibly ranging
between 21 pc and 79 pc.

We derive the space velocities, following the equations from
\citet{js1987}, and find [U, V, W]=[-16, -142, +43]\kms for the 42
pc distance. If we vary the distance, we obtain somewhat
different values, but the component of V remains consistently large ($<-90$
km/s) which rules out membership in the young disk, and again suggests
that \sdss is relatively old. The space velocities are more consistent
with inner halo kinematics \citep{cbeers} and are similar to the known
M subdwarfs \citep{lepine2003aja}. We compared the evolutionary models
of \citet{burrows2001} for a age of 10-15Gyrs, and found that for a
temperature of about 1800K for L4 spectral type \citep{kirk1999} the
initial mass is just at the hydrogen burning limit.

\section{Conclusion}

There are only three L subdwarf known at present and the
classification schemes based on colors and spectral line indexes is
difficult based on three objects. The theoretical spectrum synthesis
models are  in the preliminary stages,  due to the complicated, dust
chemistry and molecular formation at low temperatures. This makes it
difficult to formally characterize SDSS~J125637$-$0224, but evidence
strongly suggest the star should be classified as a mid-type L
subdwarf (tentatively ``sdL4''). This discovery adds up to the only
two other objects also tentatively classified as ``sdL''
\citet{burgasser2003a,burgasser2004}.

Clear morphological difference between this object and normal L dwarfs
shows it is possibility to detect such objects at low resolution.
\sdss\ also stands out in the NIR, with bluer colors than solar
metallicity L dwarfs. However since those colors are similar to field
M dwarfs, photometric identification would be a challenge without
proper motion data. Proper motion surveys in the infrared and
spectroscopic surveys in the optical and NIR of extreme red objects
would provide the best chance at identifying more objects of this kind
and establishing proper classification sequences.

\acknowledgments
TS acknowledge the support from NSF with grant NSF AST-0705139,
 NASA with grants NNX07AP14G, the UCF-UF SRI program, and 
University of Florida. 
This work is funded by the Virtual Observatory India, hosted by Inter
university Centre for Astronomy and Astrophysics. SL is supported
grant AST-0607757 from the United States National Science
Foundation. AST We would like to thank Sonali Kale and the Persistent
systems private Ltd. for customizing the VOPLOT utility to our need,
which helped in choosing the samples for this study. 

This research made use of SDSS data archive, 2MASS data archive and
SuperCOSMOS Science Archive. Funding for the creation and distribution
of the SDSS Archive has been provided by the Alfred P. Sloan
Foundation, the Participating Institutions, the NASA, the NSF, the US
DOE, the Japanese Monbukagakusho, and the Max Planck Society. The SDSS
is managed by the Astrophysical Research Consortium (ARC) for the
Participating Institutions. The Participating Institutions are the
University of Chicago, Fermilab, the Institute for Advanced Study, the
Japan Participation Group, Johns Hopkins University, Los Alamos
National Laboratory, the Max Planck Institute for Astronomy, the Max
Planck Institute for Astrophysics, New Mexico State University, the
University of Pittsburgh, Princeton University, the US Naval
Observatory, and the University of Washington. M. F. is supported in
part by the Grant-in-Aid from the Ministry of Education. 

The Two Micron All Sky Survey, is a joint project of the University of
Massachusetts and the Infrared Processing and Analysis
Center/California Institute of Technology, funded by the National
Aeronautics and Space Administration and the National Science
Foundation. The SuperCOSMOS Science Archive is prepared and hosted by
the Wide Field Astronomy Unit, Institute for Astronomy, University of
Edinburgh, which is funded by the UK Particle Physics and Astronomy
Research Council. 

This work also made use of Aladin sky server, hosted by Centre de
Donnes astronomiques de Strasbourg.

\end{document}